\documentclass[11pt]{article}
\usepackage{jheppub}
\usepackage{amsmath,epsfig}
\usepackage{subfigure}
\usepackage{amssymb,amsfonts}
\usepackage{latexsym}
\usepackage[T1]{fontenc} 
\usepackage{amsmath,amssymb,amsthm}
\usepackage{latexsym,graphicx,bbm}
 \usepackage{pdfsync}

\newcommand{\beq}{\begin{eqnarray}}
\newcommand{\eeq}{\end{eqnarray}}

\newcommand{\p}{\partial}
\newcommand{\Tr}{{\rm Tr}}
\newcommand{\tr}{{\rm tr}}

\newcommand{\sign}{{\rm sign}}

\newcommand{\ba}{\left( \begin{array}}
\newcommand{\ea}{\end{array} \right)}
\def\be{ \begin{eqnarray} }
\def\ee{ \end{eqnarray} }

\title{\boldmath Toda chain from the kink-antikink lattices}
\author[a,b]{Song He,}
\author[c,d]{Yunguo Jiang,}
\author[c]{Jiazhen Liu}

\affiliation[a]{Max Planck Institute for Gravitational Physics (Albert Einstein Institute)
Am M\"{u}hlenberg 1, 14476 Golm, Germany}
\affiliation[b]{State Key Laboratory of Theoretical Physics, Institute of Theoretical Physics, \\ Chinese Academy of Science, Beijing 100190, P. R. China}
\affiliation[c]{School of Space Science and Physics,  Shandong University at Weihai,  264209 Weihai, China}
\affiliation[d]{Shandong Provincial Key Laboratory of Optical Astronomy and Solar-Terrestrial Environment, \\ Institute of Space Sciences, Shandong University,Weihai, 264209, China}

\emailAdd{hesong17@gmail.com, jiangyg@sdu.edu.cn(Corresponding)}

\abstract{In this paper, we have studied the kink and antikink solutions in  several neutral scalar models in 1+1 dimension. We follow the standard approach to write down the leading order and the second order force between long distance separated kink and antikink. The leading order force is proportional to exponential decay with respect to the distance between the two nearest kinks or antikinks. The second order force have a similar behavior with the larger decay factor, namely $3\over 2$. We make use of these properties to construct the kink lattice. The dynamics of the kink lattice with leading order force can be identified as ordinary nonperiodic Toda lattice. Also the periodic Toda lattice can be obtained when the number of kink lattice is even.  The system of kink lattice with force up to the next order corresponds to a new specific deformation of Toda lattice system. There is no well study on this deformation in the integrable literatures.We found that the deformed Toda system are near integrable system, since the integrability are hindered by high order correction terms. Our work provides a effective theory for kink interactions and a new near or quasi integrable model.}

\begin{document}

\newpage
\pagenumbering{arabic}
\setcounter{page}{1}
\setcounter{footnote}{0}
\renewcommand{\thefootnote}{\arabic{footnote}}
\maketitle
\section{Introduction}

Kinks are soliton solutions of scalar field models in 1+1 space-time dimensions. The non-dissipative  energy density makes  kinks to be particle-like objects. The  $\varphi^4$, $\varphi^6$, $\varphi^8$ and sine-Gordon kinks are intensively studied for their interactions and scattering theories \cite{Manton:2004tk,Vachaspati:2006zz,Dorey:2011yw,Gani:2015cda}. Recently, it was found that kink solutions on the world sheet theories of non-Abelian vortices are confined monopoles in the supersymmetric QCD theories \cite{Shifman:2004dr,Eto:2011cv}. Kinks in the massive sigma model have close relation with certain spin chains \cite{AlonsoIzquierdo:2008rk,Harland:2009mf}. The dynamics of slender monopoles and antimonopoles on the vortex string can be mapped exactly onto the sine-Gordon kink system \cite{Arai:2014hda}. These work suggest us that the kink and antikink interactions can truly mimic some dynamics of integrable system.

In this paper, we focus on exploring the relationship between the multikink solutions and integrability models, especially, the Toda lattice system and its deformation. The Toda lattice is a classical integrable system introduced by Morikazu Toda, in \cite{toda}. There are many abundant contents in math and physics.
The symmetries for Toda have been developed in several articles \cite{damianou90,damianou91,damianou93}. In \cite{olper,olper1,olper2}, they discussed about various integrable systems which are related to Lie algebras. More recently,  several authors have explored how Lie group and cluster algebra are related to the Toda Lattice system \cite{Marshakov:2012kv,Fock:2014ifa,Kruglinskaya:2014pza,Fock:1997aia}. From physics point of view, the Toda system has been well studied in  \cite{damianou02,flaschka2,flaschka,henon,manakov,moser74}.

In this paper, firstly, we have constructed multikink solutions in 1+1 dimensional scalar field theory with various self-interaction potentials like $\varphi^4$, $\varphi^6$ and $\varphi^8$. After that, we study the force between kinks and antikinks with large enough distance in these theories. Our studies show that these forces seem to be universal, not only for the leading order contribution but also for the higher order corrections. At the leading order, the force is exactly coincident with the one in the  nonperiodic Toda lattice, and the number of kinks is the same as the number sites of Toda lattice. For the periodic Toda lattice, the corresponding kink lattice must have integer number of  kink and antikink pairs. This is required by the condition of connecting vacua. At the second order of the force, it is only dependent on the quadratic term of scalar potentials which seems also quite universal in cases with potential $\varphi^4$, $\varphi^6$ and $\varphi^8$. We also point out how the multikink solutions are related to the Poisson structure and cluster coordinate hidden in Toda system in terms of $2\times 2$ representation of Lax pair in Toda system. Especially, we figure out the mapping between the multikink solutions and cluster coordinate which is very helpful to study mathematical structure hidden in the multikink solution. With introducing the second order of the force, the kink lattice corresponds to the deformation of Toda lattice system with additional exponential interaction terms.  In the deformed case, we make use of Flaschka's transformation to construct the $N\times N$ representation of Lax pair. We discuss that such system is not integrable but near or quasi integrable models. The deformed periodic Toda lattice is also presented.

An overview of  the paper is as follows. In section 2, we firstly construct the kink and anti-kink solution in the 1+1 dimension scalar field with $\varphi^4, \varphi^6$ and $\varphi^8$ potentials, respectively. We make use of these solutions to calculate the leading order and second order force between the kink and antikink in these theories. In terms of the force formula, we can construct $\varphi^4, \varphi^6$ and $\varphi^8$  kink lattices, which are isomorphic to the nonperiodic Toda lattice system. In section 3, we  review various Toda lattice system and relevant Lax pair representations. We can exactly map the leading order kink lattice to the Toda lattice system. In  section 4, we take the second order force into account and claim the kink lattice with the second order force corresponds to one kind of deformed  Toda lattice. We make use of Flaschka transformations to construct the Lax pair representation of the deformed Toda lattice. The deformed Toda lattice is near integrable or quasi integrable system. In section 5, we summarize and discuss our main results in this paper. The appendix will devote to conventions in $A_N$ Lie algebra which is associated with affine $A_N$ Toda chain.

\section{The Force between Kink and Antikink   \label{sec:model}}
The force between kink and antikink can be calculated in an analytical way \cite{Manton:2004tk}.
We start with the $\varphi^4$ theory as an illustration, and show that the force has a universal formula also in the $\varphi^6$ and $\varphi^8$ case.
\subsection{$\varphi^4$ Case }
We consider the $1+1$ dimension Lagrangian for the $\varphi^4$ kink.  The Lagrangian  reads
\begin{align} \label{BL}
{\cal L}= & \frac{1}{2} \p_{\mu}\varphi \p^{\mu} \varphi - V(\varphi) ,
\end{align}
where $\mu=0,1$ and $V(\varphi)=\lambda(\varphi^2-v^2)^2$ denotes the potential. The Euler-Lagrange (EL) equation is written as
\beq \label{eq:EL}
\ddot{\varphi}-\varphi''+ \frac{d V}{d \varphi}=0.
\eeq
For the static situation, the energy of the system is given by
\beq
E=\int_{-\infty}^{+\infty}(\frac{1}{2} \varphi'^2 +V(\phi)) d x \geq \pm \int_{-\infty}^{+\infty} \sqrt{2V(\varphi)} d \varphi,
\eeq
the equal sign stands for the Bogomol'nyi bound. The BPS equation for the kink is written as
\beq
\varphi'=\pm\sqrt{2V(\varphi)} .
\eeq
The kink solution must interpolate between different vacua. Here, we define $(\varphi_{-\infty}, \varphi_{\infty})=(-v, v)$ as the kink, and  $(v, -v)$ as the antikink.
The kink solution can be written as
\beq
\varphi_1(x)= v \tanh \left[ \sqrt{2\lambda} v(x-x_{01}) \right],
\eeq
where $x_{01}$ denotes the central position of the kink. The antikink solution  can be obtained by replacing $x$ to $-x$.

For obtain the static force, we first present the definition of the momentum.
The momentum of the system can be calculated by the definition of the energy-momentum tensor, which is  written as\footnote{The  momentum tensor is defined as $T_{\mu \nu}=\frac{\delta {\cal L}}{\delta \partial \mu \phi} \partial_{\nu}\phi-g_{\mu \nu} {\cal L}$.}
\beq
P= -T_{01}= - \int_{-\infty}^b \dot{\Phi} \Phi' d x.
\eeq
Here $\Phi$ is the generic field $\varphi$ in eq.(\ref{BL}), which also depends on time.
The integration from $-\infty$ to $b$ denotes total momentum in this regime. Correspondingly, the force on the range $(-\infty, b]$ is the derived by
\beq \label{eq:force}
F=\frac{dP}{dt}=\left[-\frac{1}{2}(\dot{\Phi}^2 +\Phi'^2 )+ V(\Phi) \right]\Big|_{-\infty}^b,
\eeq
where the eq. (\ref{eq:EL}) is used. We don't consider the static condition up to this step.

Secondly, we calculate the static force between the kink and the antikink. We set  $\Phi=\varphi_1+ \varphi_2-v$, where $\varphi_1$ and $\varphi_2$ denote the static kink and  antikink solutions at position  $x_{01}$ and $x_{02}$, respectively.   We set $R= x_{02}-x_{01}> 0$ to be large but not infinite.     In order to study the static force experienced by the kink at position $x_{01}$, we set $b$ to be the center of the pair, i.e., $b = \frac{1}{2}(x_{01}+x_{02})$.
At $-\infty$, both $\varphi_1$ and $\varphi_2$ approach the vacuum, and their derivatives are zero. So, the force is related only to the pressure at $b$.
With large distance approximations $|x-x_{01}|\gg1, |x-x_{02}|\gg1$, the asymptotic formula of $\varphi_1$ and $\varphi_2$ are given by
\beq
\varphi_1\approx v \left(1-2 e^{-2\sqrt{2\lambda}v(x-x_{01})} \right), \qquad \varphi_2\approx v \left(1- 2e^{2\sqrt{2\lambda}v(x-x_{02})} \right).
\eeq
The formula are valid at the first order of $e^{-2\sqrt{2\lambda}vR}$. We set $\Delta=\varphi_2-v$, then $\Phi=\varphi_1 + \Delta$. For the static configuration, the first term $\dot{\Phi}^2$ in eq. (\ref{eq:force}) can be omitted. Then, the force $F$  can be expand according to $\Delta$, which is written as
\beq \label{fstatic}
F=\left[-\varphi_1'\Delta' +  \varphi_1''\Delta\right]\Big|_{x=b}.
\eeq

Substituting the asymptotic formula of $\varphi$ into the force $F$, we obtain that
\beq \label{1force}
F= 64v^4 \lambda e^{-2 \sqrt{2\lambda} v R}.
\eeq
The positive value of the force means that the kink at $x_{01}$ is attracted by the anti-kink at position $x_{02}$ which is consistent with the force shown in $\varphi^4$ kinks \cite{Manton:2004tk}. Inversely, the antikink must be attracted by the kink.

The procedure to obtain the force above can be applied to any kink and antikink pairs. Although the coefficients may be different for different kinds of kinks, the forces are all belong to the exponential decay pattern. We can also arrange the pairs to form the one dimensional kink lattices. We choose to use the antikink as the first site at $x_1$ in order to match the familiar Hamilton equations for Toda lattice. Secondly, we set a kink at $x_2$ on the right side of $x_1$. The distance between them is $R=x_2-x_1$. Then, we set an antikink at $x_3$ and kink at $x_4$, and so on. All of them have the equal distance $R$. Besides the first and the last one, all of them are attracted by their neighbours. For instance, we consider the force on kink at $x_2$. Its momentum is given by
 \beq
  p_2= \frac{dq_2} {dt} = \tilde{m} \dot{x}_2, \qquad
 \eeq
where $\tilde{m}=2\sqrt{2\lambda}v $. One can normalize the coefficients by setting $\lambda=1$ and $v=\frac{1}{2\sqrt{2}}$ to simplify our later analysis. Thus, one obtains
\beq
 \frac{d p_2}{dt}= -e^{q_1-q_2}+   e^{q_2-q_3}.
\eeq
which describes the force  from its neighbor kinks.
This indicates that the interaction of the kink antikink system is just the Toda lattice \cite{Manton:2004tk} at the leading order. The kink lattice can be constructed as \cite{Vachaspati:2006zz}
\beq
\Phi(x)=\sum_{i=1}^{N} \varphi_i(x-x_i) + (N_2-1) v, \qquad N_2=N \,\,\, {\rm mod}\,\,\, 2.
\eeq
where the odd $i$ denotes the kink, while the even $i$ denotes the antikink. $N$ is the number of sites of the kink lattice.

\subsection{Second Order Force}

We calculate the force by considering  only the first leading order above. The second order correction to the force is also interesting. The kink antikink configuration is expressed by $\Phi=\varphi_1+\varphi_2-v_c$, where $v_c$ is the vacuum at the center of the pair. We consider the $\Delta=\varphi_2-v_c$ as the perturbation. From eq.(\ref{eq:force}), the force expanded up to the second order is written as
\beq \label{eq:for2cor}
F=-\varphi_1' \Delta'+ \varphi_1'' \Delta -\frac{1}{2} \Delta'^2+\frac{1}{2} m_{\varphi_1}^2 \Delta^2,
\eeq
where we used the static EL equation $\Phi''=\frac{d V(\Phi)}{d \Phi}|_{\Phi=\varphi_1}$ and $m_{\varphi_1}^2=\frac{d^2 V(\Phi)}{d^2\Phi}|_{\Phi=\varphi_1}$.
The formule of $\varphi_1$ and $\Delta$ with second order correction can be approximated as
\beq  \label{eq:phidel2exp}
\varphi_1 \approx v(1-2\delta_1+2\delta_1^2), \qquad \Delta \approx -2v \delta_2+2v\delta_2^2,
\eeq
where $\delta_1=e^{-\tilde{m}(x-x_1)}$ and $\delta_2=e^{\tilde{m}(x-x_2)}$.  One can check easily that $m_{\varphi_1} \approx \tilde{m}$. Substituting eq.(\ref{eq:phidel2exp}) into $F$ in (\ref{eq:for2cor}), one obtains that
\beq\label{secondorder}
F= 8\tilde{m}^2v^2 \left[ e^{-\tilde{m}R}-4 e^{-\frac{3}{2}\tilde{m}R}+\cdots\right].
\eeq
where dots denote the higher order corrections. The second order correction is the second term above, which means the corrected force is attractive. With the normalization $\tilde{m}=1$ and $v=\frac{1}{2\sqrt{2}}$, one obtains $F=e^{-R}-4 e^{-\frac{3}{2}R}$.
Along the kink lattice, the second soliton experiences the following force
\beq
\frac{dp_2}{dt}= -e^{q_1-q_2}+   e^{q_2-q_3}+4e^{\frac{3}{2}(q_1-q_2)}-4e^{\frac{3}{2}(q_2-q_3)}.
\eeq
One can write all the forces experienced by the solitons in the subsequent lattice.
\subsection{$\varphi^6$ Case}
The Lagrangian for the $\varphi^6$ model reads \cite{Dorey:2011yw,Gani:2014gxa}
\beq
{\cal L}=\frac{1}{2}\p_{\mu}\varphi\p^{\mu}\varphi -\frac{1}{2}\varphi^2(\varphi^2-1)^2.
\eeq
 Three vacua exist for constructing the kink solution. For the boundary $(-\infty,\infty)=(0,1)$, the kink solution is given by $\varphi_{(0,1)}=\varphi_1=\sqrt{(1+\tanh x)/2}$. Its antikink $\varphi_{(1,0)}=\varphi_2$ is obtained by the reflection $x \to -x$. Now we arrange the kink and antikink at $x_{01}$ and $x_{02}$ positions, respectively. The inter force between them can be calculated following the procedure above. First, the asymptotic formula of $\varphi_{1}$ and $\varphi_{2}$ at $x=(x_{01}+x_{02})/2$ are given by
 \beq
 \varphi_{1}\approx 1-\frac{1}{2}e^{-2(x-x_1)}, \qquad  \varphi_{2}\approx 1-\frac{1}{2}e^{2(x-x_2)},
 \eeq
 respectively. The internal force between them is calculated to be
 \beq
 F=[-\varphi_1'\varphi_2' +\varphi_1'' (1+\varphi_2)]\Big|^{\frac{x_1+x_2}{2}}_{-\infty}\approx2 e^{-2R}.
 \eeq
This force agrees with the normalized force in the $\varphi^4$ case. The inter forces between kink and its antikink  with different vacua can also be obtained in the same way. The second order correction to the force can be calculated similarly, which is expressed as
\beq
F = 2 \left[ e^{-2R}-4e^{-3R}+ \cdots\right],
\eeq
where higher order corrections are omitted. Note that the coefficients of the second order agrees with the $\varphi^4$ case.

\subsection{$\varphi^8$ Case}
Now we consider the $\varphi^8$ kink case \cite{Gani:2015cda}. The Lagrangian is given by
\beq
{\cal L}=\frac{1}{2}\p_{\mu}\varphi\p^{\mu}\varphi -\lambda^2(\varphi^2-v_1^2)^2(\varphi^2-v_2^2)^2.
\eeq
Four vacua are given by $\pm v_1,\pm v_2$. First, we consider the force between $\varphi_{(-v_1,v_1)}=\varphi_1$ and $\varphi_{(v_1,-v_1)}=\varphi_{2}$. The kink solution satisfies
\beq
\frac{v_1-\varphi_{a}}{v_2+\varphi_a}=e^{-\mu x} \left( \frac{v_2-\varphi_a}{v_2+\varphi_a}\right)^{v_1/v_2},
\eeq
where $\mu=2\sqrt{2\lambda}v_1(v_2-v_1)^2$.
Setting $\varphi_1$ and $\varphi_2$ at position $x_{01}$ and $x_{02}$, respectively. At the center of them, the asymptotic formule are written as
\beq
\varphi_1 \approx v_1-2v_1 \left( \frac{v_2-v_1}{v_1+v_2}\right)^{\frac{v_1}{v_2}} e^{-\mu (x-x_{01})}, \qquad \varphi_2 \approx v_1-2v_1 \left( \frac{v_2-v_1}{v_1+v_2}\right)^{\frac{v_1}{v_2}} e^{\mu (x-x_{02})} ,
\eeq
respectively. The inter force between $\varphi_1$ and $\varphi_2$ is calculated to be
\beq \label{eq:phi8v1}
F=8v_1^2\mu^2 \left( \frac{v_2-v_1}{v_1+v_2}\right)^{2\frac{v_1}{v_2}} e^{-\mu R}.
\eeq
Consider the second order correction, the force becomes
\beq
F=8v_1^2\mu^2 \left( \frac{v_2-v_1}{v_1+v_2}\right)^{2\frac{v_1}{v_2}} \left[ e^{-\mu R} -4e^{-\frac{3}{2}\mu R}+ \cdots  \right].
\eeq

Secondly, we consider another kind of $\varphi^8$ kink solution. The kink interpolating between $-v_1$ and $-v_2$ is denoted as $\varphi_3$. Its solution satisfies
\beq
e^{\mu x}=\frac{\varphi_3-v_1}{\varphi_3+v_1}\left( \frac{v_2+\varphi_3}{v_2-\varphi_3}\right)^{\frac{v_1}{v_2}}.
\eeq
Consider $\varphi_3$ at position $x_{03}$, and its anti company $\varphi_4$ at position $x_{04}$, respectively. The inter force between them is calculated to be
\beq  \label{eq:phi8v1v2}
F=8v_1^2 \mu^2 \left( \frac{v_2 -v_1}{v_2+ v_1}\right)^{\frac{2v_1}{v_2}} e^{-\mu R}.
\eeq
We notice that this force is the same with that between $\varphi_1$ and $\varphi_2$ in Equation (\ref{eq:phi8v1}).

Now we consider the inter force between two different kinds of kink. Suppose that a kink of $\varphi_{(-v_2,-v_1)}=\varphi_5$ is at position $x_{01}$, while a kink of $\varphi_{(-v_1,v_1)}=\varphi_{06}$ sits at $x_{02}$. These multiple kink solution can be written as $\varphi=\varphi_5+\varphi_6+v_1$. The asymptotic form of $\varphi_5$ and $\varphi_6$ are given by
\begin{align}
\varphi_5 \approx & -v_1-2v_1 \left( \frac{v_2-v_1}{v_2+ v_1}\right)^{\frac{v_1}{v_2}} e^{-\mu(x-x_{01})}, \qquad {\rm for }\qquad x \gg x_1, \\
\varphi_6 \approx & -v_1+2v_1 \left( \frac{v_2-v_1}{v_2+ v_1}\right)^{\frac{v_1}{v_2}} e^{\mu(x-x_{02})}, \qquad {\rm for }  \qquad x_2 \ll x_1,
\end{align}
respectively. The inter force between them are given by
\beq
F=-8v_1^2 \mu^2 e^{-\mu R}.
\eeq
The minus sign of this force means that the $\varphi_5$ kink experiences an expelling effect from $\varphi_6$. This is quite different with the force between kink and antikink, which is always attractive. We can also take use of $\varphi_5$ and $\varphi_6$ kinks to construct the kink lattice. The corresponding potential in the Hamiltonian only changes a sign. However, the vacua of different kinks must be correctly connected along the kink lattice. More kinds of kinks interaction will lead to the inhomogeneous Toda lattice, which will be studied in future.

To close this section, we comment on the leading order and second order force between kinks. We can find that the leading and second order forces are not sensitive to the scalar potential in 1+1 dimensional field theory with $\varphi^4,\varphi^6$ and $\varphi^8$ scalar potentials. The force just only involves in the mass of the scalar, and the higher power scalar self-interactions do not make contributions to the force in terms of current study. One main reason is that we just take use of large distance between kinks as approximation to obtain the force. In this approximation, the mass term will be dominant in contributing to the force.
Further, the second order correction to the force is proportional to $e^{-{3\over 2}m|\text{distance}|}$ and has a reverse direction against the leading order. Some resonance phenomena in kink collisions may be caused by the high order corrections. One can see that factor ${3\over 2}$ is quite independent on the scalar potential. In the second order force, one more thing is that the coefficient of $e^{-{3\over 2}m|\text{distance}|}$ will depends on the potential parameter.
Other kink theories can be tested to see whether the coefficients of high order corrections are universal or not.
Finally, we can construct long distance kink lattice and identify this lattice to be the nonperiodic Toda lattice. We can also find the new classical near  integrable model which corresponds to one specific deformation of Toda spin chain system.

\section{Toda lattice system \label{sec:TC}}

Once we have the force between the kink and antikink, we can construct the kink lattices to represent the Toda lattice. The kink and antikink alternately appear on a string in one direction, which forms the kink lattice. For ${\cal A}_N^{(1)}$ Toda lattices \cite{Bo}, the Hamiltonian is written as
\beq \label{Todachain}
H=\sum_{j=1}^N\frac{p_j^2}{2}+V(q)\,,\qquad \{p_j,q_k\}=\delta_{jk}\,,
\eeq
where
\beq
V(q)&=&V_{{\cal A}_N}+\exp(q_N-q_1)\,, \qquad V_{{\cal A}_N}=\sum_{j=1}^{N-1}\exp(q_j-q_{j+1})\,.
\eeq
The periodic ${\cal A}_N^{(1)}$ Toda lattices which corresponds to the root systems of affine algebras \cite{Bo}.

The equations of motion of the  periodic Toda lattice with $N$  sites reads \cite{GKMMM,MW,6,61}
\beq\label{Todaeq1}
\frac{\partial q_i}{\partial t} &=& p_i, \nonumber\\
\frac{\partial p_i}{\partial t} &=& e^{q_{i+1} -q_i}-
e^{q_i-q_{i-1}},
\eeq
where $q_i$ and  $p_i$ denote the position  and momentum for each lattice, respectively.
The periodic boundary conditions for lattice are $p_{i+N}=p_i$, $q_{i+N}=q_i$.
There are exactly $N$ conservation laws in the $SU(N)$ Toda lattice system.
These conservation quantities can be exactly constructed from the Lax operator.

In this section, we would like to present two different Lax representations to describe the periodic Toda lattice. Basing on these representations,
one can find the dictionary between the Toda lattice and the kink lattice system.
The Lax operator has been well studied in the integrable literature.
In the first representation, the Lax operator is given by the $N\times N$
matrix depending on dynamical variables $q_i$ and $p_j$,
\beq\label{LaxTC1}
{\cal L}(w) =
\left(\begin{array}{ccccc}
p_1 & \text{ } e^{{1\over 2}(q_2-q_1)} & 0 & & we^{{1\over 2}(q_1-q_{N})}\\
e^{{1\over 2}(q_2-q_1)} & p_2 & e^{{1\over 2}(q_3 - q_2)} & \ldots & 0\\
0 & e^{{1\over 2}(q_3-q_2)} & p_3 & & 0 \\
 & & \text{ }  \ldots & & \\
\frac{1}{w}e^{{1\over 2}(q_1-q_{N})} & 0 & 0 & & p_{N}
\end{array} \right).
\eeq
The characteristic equation for the Lax matrix can be defined as
\beq\label{SpeC1}
{\cal P}(\lambda,w) = \det_{N\times
N}\left({\cal L}(w) - \lambda\right) = 0,
\eeq
and the corresponding spectral curve is written as following
\beq\label{fsc-Toda1}
w + \frac{1}{w} = 2P_{N}(\lambda ),
\eeq
where $P_{N}(\lambda)$ is a polynomial of degree $N$.
The spectral curve (\ref{fsc-Toda1}) can be also put into the following form
\beq\label{21}
2Y\equiv w-{1\over w},\ \ \ Y^2=P_{N}^2(\lambda)-1.
\eeq

The second Lax representation appeals to introduce the
local operators at each site of the chain
\beq\label{LTC11}
L_i(\lambda) =
\left(\begin{array}{cc} \lambda -p_i & e^{q_i} \\ -e^{-q_i} & 0
\end{array}\right), \ \ \text{ } \ i = 1,\dots ,N.
\eeq
This Lax operator makes a shift to the neighbour site and transform to be
\beq\label{lp21}
L_i(\lambda)\Psi_i(\lambda)=\Psi_{i+1}(\lambda),\text{ }
\eeq
where $\Psi_i(\lambda)$ is often called by the two-component Baker-Akhiezer function.
The periodic boundary conditions can be formulated as
\beq\label{TCbc1}
\text{ }  \Psi_{i+N}(\lambda)= w\Psi_i(\lambda),
\eeq
where $w$ is a free parameter. One can also introduce the
transfer matrix that shifts $i$ to $i+N$, i.e.,
\beq\label{monomat1}
\text{ }  T(\lambda)\equiv L_{N}(\lambda)\ldots L_1(\lambda).
\eeq
The periodic boundary conditions indicate that $T(\lambda)\Psi_i(\lambda)=
w\Psi_i(\lambda)$. Thus, one has
\beq\label{specTC01}
\text{ } \det_{2\times 2}\left( T_{N}(\lambda )- w\right) &=& 0.
\eeq
The spectrum curve in terms of second representation of Lax operator is the same as the one in first representation of lax operator
\begin{align}\label{spectrum}
\text{ } 0=&w^2 - w\Tr T_{N}(\lambda ) + \det
T_{N}(\lambda ) = w^2 - w\Tr T_{N}(\lambda ) + 1,\nonumber\\
\ \ \ \ \ 0 =& \Tr T_{N}(\lambda) - w -
\frac{1}{w} = 2P_{N}(\lambda) - w - \frac{1}{w} = {\cal P}(\lambda ,w).
\end{align}
That also means the two different Lax representations are equivalent.

In the previous section, we have construct the kink lattice and we find that the leading order interaction potential between neighborhood two lattices  in large distance approximation is the same as the Toda potential in 1+1 scalar field theory.
Therefore, we can directly use the kink lattice system to construct the  $SU(N)$ Toda chain system approximately. The positions and momentums of the kinks and antikinks  can be directly mapped to the dynamical variables $q_i$ and $p_i$, respectively. Both the periodic and nonperiodic Toda chains can be constructed. All details are given in section 5.


In \cite{Marshakov:2012kv} \cite{Fock:2014ifa}, the author had discussed the Poisson structures on Lie groups $A_N$ and proposed an obvious construction of the integrable models on their corresponding Poisson submanifolds. Recently, \cite{Kruglinskaya:2014pza} extended the construction of the relativistic Toda chains as integrable systems on the Poisson submanifolds in Lie groups beyond the case of the $A$ series. They introduce the cluster structure on the Poisson submanifolds in Lie groups, which should be related to our building block $2\times 2$ representation of periodic $A_1$ Toda chain. The main motivation here is to review that the modular parameters can be interpreted as the cluster variables or Darboux coordinates in corresponding Poisson submanifolds in Lie group $A_1$ explicitly.

In the simply-laced Lie group $G$, the Poisson quivers of the Toda symplectic leaves in $G/\mbox{Ad}H$ can be drawn just by ''blowup'' of the corresponding Dynkin diagrams. In \cite{Marshakov:2012kv}, the coordinates on the corresponding leaf and the Poisson structure is defined in this coordinates by
\begin{align}
\label{pbN1}
\{ y_i,x_j\}_{\mathfrak{g}} &= C_{ij}^{\mathfrak{g}}y_ix_j,\ \ \ i,j=1,\ldots, \mbox{rank} G=N-1\nonumber
\\
\{ x_i,x_j\}_{\mathfrak{g}}& = 0,\ \ \ \{ y_i,y_j\}_{\mathfrak{g}} = 0,
\end{align}
for $\mathfrak{g}=A_{N-1}$, where $C_{ij}^{\mathfrak{g}}$ is the corresponding Cartan matrix (e.g. for $A_{N-1}$ it can be explicitly written in the form of (\ref{CslN1})).

For loop group of $A_{N-1}$, authors \cite{Marshakov:2012kv} shown that  the Lax pair representation is defined through the following way
\be
\label{notloop1}
{\bf H}_i(z) = H_i(z)T_z,\ \ \text{ }
\mathbf{E}_i =E_i,\ \ \text{ }  \mathbf{E}_{\bar i} =E_i^{\rm tr}.
\ee
where (one can see the Appendix~\ref{ap:promunu1})
\be\label{shiftT1}
E_i&=&\exp (e_i),\ \ \text{ }  \ E_{\bar i}=\exp (e_{\bar i}),\ \ \text{ }  \ H_i(z) = \exp(h^i\log z),\nonumber\\
T_z &=& \text{ }  \exp\left({\log z\frac{\partial}{\partial \log\lambda}}\right) = z^{\lambda\partial/\partial \lambda}.
\ee
For generic $\widehat{SL(N)}$ Toda system, the $2\times 2$ representation of Lax pair \cite{Marshakov:2012kv} can be expressed by
\beq
\label{prMN1}
\text{ } L=\Xi(x_0,y_1)\Xi(x_1,y_2)\ldots\Xi(x_{N-1},y_N).
\eeq
 Here
\beq
\label{xiop1}
\Xi(x,y) &=& \mathbf{H}_0(x)\mathbf{E}_0\omega \mathbf{E}_{\bar 0}\mathbf{H}_1(y) \nonumber
\\
&=& T_x\cdot \left(
\begin{array}{cc}
   1 & 0 \\
 \lambda  & 1
\end{array}\right)\left(
\begin{array}{cc}
   0 & \lambda^{-1/2} \\
 \lambda^{1/2} & 0
\end{array}\right)\left(
\begin{array}{cc}
   1 & 1/\lambda \\
   0 & 1
\end{array}\right)\left(
\begin{array}{cc}
  y^{1/2} & 0 \\
  0 & y^{-1/2}
\end{array}\right)\cdot T_y.
\eeq
One can introduce  transformation variables to rewrite eq.(\ref{prMN1}), i.e.,
\be
\label{xyxe1}
x_i = {\xi_i\over\xi_{i+1}},\ \ \ y_i = {\eta_i\over\eta_{i+1}},\ \ \
\xi_N = \xi_0,\ \ \ \eta_N = \eta_0,\ \ \ i=1,\ldots,N-1.\nonumber
\ee
So, the Poisson brackets in eq.(\ref{pbN1}) can be expressed by
\be
\label{xietaN}
\{ \eta_i, \xi_j\} = \delta_{ij}\eta_i\xi_j,\ \ \ \ i, j=1,\ldots,N.
\ee

The coordinates and momenta $\{ q_i,p_j\} = \delta_{ij}$ for eq.(\ref{xietaN}) are introduced by
\beq
\label{xew1}
\xi_i &=& \exp(-q_i),\nonumber\\
\eta_i & =& \exp(P_i+q_i),\ \ \ i=1,\ldots,N\nonumber
\\
P_i&=&p_i+{\partial\over\partial q_i}\left( {1\over 2}\sum_{k=1}^N
{\rm Li}_2\left(-\exp(q_k-q_{k+1})\right)\right).
\eeq
The $\Xi$-operators in (\ref{prMN1}) can be rewritten as
\beq
\label{prMNL1}
\Xi(x_0,y_1)\Xi(x_1,y_2)\ldots\Xi(x_{N-1},y_N) \simeq \prod_{j=1}^N L_j(\eta_j,\xi_j;\lambda )
= {\cal T}_N(\lambda )
\eeq

One can get the product of the matrices
\beq
\label{L22}
L_j(\lambda )
= \left(\begin{array}{cc}
                      0 & \sqrt{\xi_j\over \lambda \eta_j} \\
                      \sqrt{\lambda \eta_j\over \xi_j}\ \ & \ \sqrt{\lambda \over \eta_j\xi_j}+\sqrt{\eta_j\xi_j\over \lambda }
                    \end{array}\right)=
\left(\begin{array}{cc}
                      0 & {e^{-P_j/2-q_j}\over \sqrt{\lambda }} \\
                      \sqrt{\lambda }e^{P_j/2+q_j}\ \ & \ \sqrt{\lambda }e^{-P_j/2}+{e^{P_j/2}\over \sqrt{\lambda }}
                    \end{array}\right).
                    \\
j=1,\ldots,N \nonumber
\eeq

In the context of $2\times 2$ formalism, the Lax matrix (\ref{L22}) turns the algebraic limit (roughly speaking: linear in momenta, exponentiated co-ordinates and $\zeta$, as well as $\lambda  =i\exp(\zeta/2)$), into
\beq
\label{11L22nr}
{1\over i}\left(\begin{array}{cc}
                      0 & {e^{-P_j/2-q_j}\over \sqrt{\lambda  }} \\
                      \sqrt{\lambda  }e^{P_j/2+q_j}\ \ & \ \sqrt{\lambda  }e^{-P_j/2}+{e^{P_j/2}\over \sqrt{\lambda  }}
                    \end{array}\right)\rightarrow
                    L_j(\zeta) = \left(\begin{array}{cc}
                      0 & -\exp\left(-q_j/2\right) \\
                      \exp\left(q_j/2\right)\ \ & \ \zeta-p_j
                    \end{array}\right),
                    \\
j=1,\ldots,N \nonumber
\eeq
From the (\ref{11L22nr}), we can obtain the $2\times 2$ representation eq.(\ref{lp21}) of Lax pair representation in the algebraic limit \cite{Marshakov:2012kv}. In this subsection, one can easily see how these dynamical quantities $p_i$ and $q_j$ in Toda system are associated with cluster coordinates or Darboux coordinates explicitly. One can also obtain $p_i$ and $q_j$ in multiple kink solution can be map to cluster coordinates which are associated with Poisson structure. Here, we just find the coincident and it will be nice to explore the more deeper understanding in our future work.

\section{Flaschka's Transformation for {$A_N$} Toda Chain}

In this section, we will review Flaschka's transformation in Toda lattice. We will make use of this transformation to study one kind of deformation of Toda lattice. The equation of motion for $A_N$ Toda lattice is described by the equation (\ref{Todachain}). In this system, the symplectic bracket defined in $\mathbb{R}^{2N}$ is given by
\beq
\{ f, g \}_s = \sum_{i=1}^N \left( {\partial f \over \partial q_i} {\partial g
\over \partial p_i} - {\partial f \over \partial p_i} { \partial g \over
\partial q_i} \right).
\label{Symplectic_bracket1}
\eeq

One can make use of Flaschka's transformation \footnote{This is an efficient simple way to obtain the Lax pair representation. In this paper, we make use of Flaschka's transformation to construct the Lax pair representation of the deformed Toda chain system.} to obtain the $N\times N$ Lax pair representation
\begin{gather}
\begin{split}
a_i &= {1 \over 2} e^{ {1 \over 2} (q_i - q_{i+1} ) }, \,\, i=1,2,\ldots, N-1\\
b_i& =-{ 1 \over 2} p_i, \,\, i=1,2,\ldots,N.
\label{eqFab11}
\end{split}
\end{gather}
The dynamical system (\ref{Todaeq1}) transforms to be
\be
\begin{array}{lcl}
\dot a _i& = & a_i \,  (b_{i+1} -b_i ), \,\,\, i=1,2,\ldots, N-1    \\
\dot b _i &= & 2 \, ( a_i^2 - a_{i-1}^2 ), \,\,\, i=1,2,\ldots, N.
\label{Toda_lattice}
\end{array}
\ee

Flaschka's transformation is very helpful to find integrals of motion for the Toda lattice.
It can be easily verified that equations (\ref{Toda_lattice}) are
equivalent to the Lax equation $\dot{L}=[B,L]$ where
\be
L=
\begin{pmatrix} \label{1eqlaxL11}
b_1 & a_1 & 0 & \ldots & 0 & 0\\
a_1 & b_2 & a_2 &   & 0 & 0\\
0 & a_2 & b_3 & \ddots & \vdots & \vdots \\
\vdots & \vdots & \ddots & \ddots & \ddots & \vdots \\
0 & 0 & 0 & \ddots & b_{N-1} & a_{N-1} \\
0 & 0 & 0 & \cdots & a_{N-1} & b_N
\end{pmatrix},
\ee
and the $B$ is corresponding to the skew-symmetric part of $L$, which is given by
\be
B=
\begin{pmatrix} \label{1eqLaxB111}
0 & a_1 & 0 & \ldots & 0 & 0\\
-a_1 & 0 & a_2 &   & 0 & 0\\
0 & -a_2 & 0 & \ddots & \vdots & \vdots \\
\vdots & \vdots & \ddots & \ddots & \ddots & \vdots \\
0 & 0 & 0 & \ddots & 0 & a_{N-1} \\
0 & 0 & 0 & \cdots & -a_{N-1} & 0
\end{pmatrix}.
\ee
The Lax pair $L$ has the property that its eigenvalues are invariant over time
and therefore the conserved charges $H_i=\frac{1}{i}\tr{L^i}$ are constants of motion for the Toda lattice.

There exists a bracket in Flaschka's coordinates which comes from   equations
\begin{align}
\{a_i,b_i\} =& -a_i, \qquad i=1,2,\ldots,N,\nonumber\\
\{a_i,b_{i+1}\} =& a_i, \qquad i=1,2,\ldots,N-1,
\end{align}
and all other brackets are zero. One should note that we here imposed periodic boundary condition mentioned above.
The function $H_1=b_1+b_2 + \dots +b_N$ is the only Casimir for this bracket. The Hamiltonian
$H_2 = {1 \over 2} {\rm tr}  L^2=\frac{1}{2}(b_1^2+\ldots+b_N^2)+a_1^2+\ldots+a_{N-1}^2
$ gives equations (\ref{Toda_lattice}).
The functions $H_2,\ldots, H_N$ ensure the integrability of the system.
They are independent and in involution (i.e., $ \{  H_i, H_j \}=0$).

\section{Lax Pair for Deformed Toda Chain}

In this section we will present the Lax pairs for both the nonperiodic and periodic deformed Toda lattice, respectively. First, we present the nonperiodic case. The Hamiltonian for $N$ sites of kink lattice can be written as
\beq
H=\frac{1}{2}\sum_{i=1}^{N}p_i^2 +\sum_{i=1}^{N-1}\left(- e^{q_i-q_{i+1}}+\frac{8}{3}e^{\frac{3}{2}(q_i-q_{i+1})}\right).
\eeq
where the second term in the summation is corresponding to deformation of Toda system.
The dynamics of the $i$th site is given by
\begin{align}
\dot{q}_i=&p_i, \\
\dot{p}_1=&e^{q_{1}-q_2}-4e^{\frac{3}{2}(q_{1}-q_2)} , \qquad \dot{p}_N=-e^{q_{N-1}-q_N}+4e^{\frac{3}{2}(q_{N-1}-q_N)},\\
\dot{p}_i=&-e^{q_{i-1}-q_i}+ e^{q_i-q_{i+1}}+4e^{\frac{3}{2}(q_{i-1}-q_i)}-4e^{\frac{3}{2}(q_i-q_{i+1})}, \quad i=2,\cdots,N-1.
\end{align}
Now we use the Flaschka's transformation to construct the Lax pair.
First, we assume that the $L$ matrix still has the fomula in eq.(\ref{1eqlaxL11}). Since the deformed dynamics should go back to the Toda chain dynamics without the second order force. However, the expression of $a_i$ and $b_i$ should be changed. The conserved quantities are given by $H_i=\frac{1}{i}L^i$. For $i=1$ and $i=2$, the momentum and energy of the system are conserved. So, we choose $b_i=p_i$ to satisfy the conserved law.
Then, ansatz for $a_i$ needs to be given. From eq.(\ref{eqFab11}), we guess that
\beq  \label{eqa}
a_i=&\sqrt{-e^{q_i-q_{i+1}}+\frac{8}{3}e^{\frac{3}{2}(q_i-q_{i+1})}},
\eeq
which agrees with Toda chain system if the high order corrected force is not present. Note that we have add a minus sign in the potential term.
Then, $B$ matrix should have such a property $B^{T}=-B$, namely antisymmetric. We further assume that $B$ is given by
\begin{align} \label{eq:B}
B=\left(\begin{array}{ccccc}
0 & c_1 & 0 & \ldots & 0 \\
-c_1 & 0 & c_2 &  & \vdots \\
0 & -c_2 & 0 & \ddots & \vdots \\
\vdots &  & \ddots & 0 & c_{N-1} \\
0 & \ldots & \ldots & -c_{N-1} & 0 \\
\end{array} \right).
\end{align}
Then the Lax pair equation $\dot{L}=[B,L]$ gives us such relations
\begin{align}
\dot{a_i}=&c_i(b_{i+1}-b_i), \qquad i=1,\cdots,N-1 \\
\dot{b}_1=&2c_1a_1, \\
\dot{b_i}=&2(c_ia_i-c_{i-1}a_{i-1}), \qquad i=2, \cdots, N-1 \\
\dot{b}_N=&-2a_{N-1}c_{N-1}.
\end{align}
And the solution of $c_i$ is given by
\beq
 c_i=\frac{\partial a_i}{\partial q_i}.  \qquad i=1,\cdots,N-1
\eeq
We find that the validity of the Lax pair needs the constraint that
$c_{i+1} a_i- c_{i}a_{i+1}$, which is the $(i,i+2)$ component of $[M,L]$, should be zero. If we keeping only the leading order interaction, this constraint
is satisfied. However, the constraint are not satisfied when high order exponential terms are included. The component $(c_i a_{i+1}-a_i c_{i+1})$ are high order
interaction correction. This indicates that the kink lattice is not exact but "near" integrable. For the only two sites case, the deformed Toda theory are exact integrable, since there are no constraint condition anymore.
The kink lattice with more high order corrections are studied in another work in preparation \cite{Jiang:2016a}.

The conserved quantities is obtained as
\beq
H_i=\frac{1}{i} {\rm Tr} L^i, i=1, \cdots, N
\eeq
We give several examples for illustration. For $N=2$ case, we have the momentum and energy conservation laws, i.e.,
\begin{align}
H_1=&p_1+p_2,\\
H_2=&\frac{1}{2}p_1^2+\frac{1}{2}p_2^2-e^{q_1-q_{2}}+\frac{8}{3}e^{\frac{3}{2}(q_1-q_{2})}.
\end{align}
$H_2$ is exact the Hamiltonian.
The $N=3$ case is the first "near" integrable case, we would like to study such system for illustration. Although this case is not exact integral, it deserves to be studied. One can obtain three integrals of motion  in the following
\begin{align}
H_1=&p_1+p_2+p_3,\\
H_2=&\frac{1}{2}(p_1^2+p_2^2+p_3^2)+a_1^2+a_2^2,\\
H_3=&\frac{1}{3}(p_1^3+p_2^3+p_3^3)+(p_1+p_2)a_1^2+(p_2+p_3)a_2^2.
\end{align}
where $a_1$ and $a_2$ are expressed in eq.(\ref{eqa}).  We also calculate the time derivative of the third conserved quantity, and find that
\beq
\{ H_1, H_2\}=0, \qquad \{H_1,H_3 \}=0, \qquad \{H_2,H_3 \}=2a_1a_2(a_1c_2-a_2c_1).
\eeq
The last equation indicates us that $H_3$ is not a conserved quantity. This means that there is perturbation for the "near" integrable system.

Secondly, we construct the periodic deformed Toda lattice from the kink lattice. The Hamiltonian for the kink lattice  can be given by
\beq
H=\frac{1}{2}\sum_{i=1}^{N}p_i^2 +\sum_{i=1}^{N-1}\left(- e^{q_i-q_{i+1}}+\frac{8}{3}e^{\frac{3}{2}(q_i-q_{i+1})}\right)- e^{q_N-q_{1}}+\frac{8}{3}e^{\frac{3}{2}(q_N-q_{1})}.
\eeq
In order to obtain the interaction between the first and the $N$th site, we must glue the head and the tail of the kink lattice with $N$ sites. In order to
correctly connect them, the left vacuum of the first site must agrees with the right vacuum of the $N$th site. Thus, only kink and antikink can be connected.
This put a constraint that the number of sites for periodic kink lattice must be an even integer. Thus, the kink lattice can only represent periodic  Toda lattice with even number sites.

The Lax matrix for the periodic deformed Toda lattice can be written as
\beq
L(\omega)=
\begin{pmatrix} \label{1eqlax2}
b_1 & a_1 & 0 & \ldots & 0 & \omega a_0\\
a_1 & b_2 & a_2 &   & 0 & 0\\
0 & a_2 & b_3 & \ddots & \vdots & \vdots \\
\vdots & \vdots & \ddots & \ddots & \ddots & \vdots \\
0 & 0 & 0 & \ddots & b_{N-1} & a_{N-1} \\
\frac{1}{\omega}a_0 & 0 & 0 & \cdots & a_{N-1} & b_N
\end{pmatrix},
\eeq
where $a_0=\sqrt{-e^{q_N-q_{1}}+\frac{8}{3}e^{\frac{3}{2}(q_N-q_{1})}}$. Other components in $L$ and $B$ matrices are the same as the open Toda lattice case. Besides, the periodic conditions are $q_{i+N}=q_i$ and $p_{i+N}=p_i$.
 The characteristic equation for the Lax matrix can be calculated following the procedure in Sec. \ref{sec:TC}.

\section{Conclusions and Discussions}

In this paper, we have studied the kink and antikink solutions in 1+1 scalar field theories with various scalar potentials. In these theories, we figure out the general formula for the force between kinks and antikinks. The force shows some universal property up to the  second order interactions. The leading order of the force is proportional to $e^{-|q_i-q_{i+1}|}$, where $q_i$ are positions of the kinks in the spatial direction. Motivated by the exponential form of the force, we  construct the kink lattice, in which the nearest neighborhood positions are always occupied by a kink and antikink. If the distance between neighbors is large enough, the kink lattice  can be exactly regarded as the  $N$ sites Toda lattice system. Both periodic and nonperiodic Toda lattices can be constructed. For periodic kink lattice, there should be pairs of kink and antikink on the lattice. The boundary conditions are the same for both the kink and the Toda lattices. We  make use of $2\times 2$ representation of Lax pair in Toda chain system to find the connection between $q_i, p_i$ in the kink lattice and cluster coordinates in Toda chain system.

Further, we  have studied the second order interaction between kink and antikinks, which are shown  to be proportional to $4 e^{-{3\over 2}(|q_i-q_{i+1}|)}$. The different direction of the first and the second order forces may be related to the resonance phenomena in kink scattering tests.
The kink lattice with  the second order interaction must correspond to the deformed Toda chain system.  To study its integrability, one should construct the Lax pair representation of the deformed Toda system, or find the finite  conserved charges. By using the Flaschka's transformation, we find a Lax representation for the model. However, the high order correction term ruin the integrability of the system. Thus, the kink lattice is not an integrable but an near or quasi integrable system \cite{Ferreira:2010gh,Ferreira:2016ubj}. Since our kink solutions are based the $1+1$ dimension theories, which can be studied by the quasi-integrable analysis method.    Our study may also be applied to domain walls in literatures directly \cite{Gani:2016txo}. In SQCD, the kink solutions on the non-Abelian vortex string are the confined monopoles \cite{Shifman:2004dr,Eto:2011cv,Gudnason:2010rm}. These kinks are solutions of massive sigma model, which is the low energy theory of the non-Abelian vortices. The static forces between them can be repulsive \cite{Arai:2014hda}.  One can expect  a non-homogeneous deformed Toda chain from the generic massive sigma model. The inter force between non-Abelian kinks may have contributions from the orientational modes \cite{Eto:2015uqa,Nitta:2014rxa}. The Toda system from such non-Abelian solitons deserve future studies.   In summary, the kink lattice is a well representation of the deformed Toda chain system. The correspondence between them will enable us to explore their properties in a reciprocal manner.

\vskip 0.5cm
{\bf Acknowledgement}
\vskip 0.2cm
We are grateful to Dmitri Bykov, Vakhid A. Gani, Yutin Huang, Dharmesh Jain, Andrei Mironov, Axel Kleinschmidt, Muneto Nitta, Tadashi Takayanagi, Stefan Theisen, Kentaroh Yoshida for useful conversations and correspondence. S.H. is supported by Max-Planck fellowship in Germany and the National Natural Science Foundation of China (No.11305235). The work of Y. G. Jiang is supported by Shandong Provincial Natural Science Foundation (No.ZR2014AQ007) and National Natural Science Foundation of China (No. 11403015).

\appendix
\renewcommand{\theequation}{\thesection.\arabic{equation}}
\addcontentsline{toc}{section}{Appendices}

\section{Notations for Lie Algebra
\label{ap:promunu1}}

For given a Cartan matrix $C_{ij}$, the Lie algebra $\mathfrak{g}$ is generated by $\{h_i|i \in \Pi\}$ and $\{e_i|i \in \Pi\cup{\bar\Pi}\}$, which are labeled by simple positive $\Pi$ and negative $\bar{\Pi}$ roots~\footnote{In order to simplify their notation, we extend $h$ and $C$ to negative values of indices with assuming  $h_i = h_{-i}$ and $C_{-i,-j}=C_{ij}$ and $C_{ij}=0$ if $i$ and $j$ have different signs. We also use the notation $e_{-i}=e_{\bar i}$ for $i>0$.}. They satisfy with following the commutation relations
\beq\label{rela1}
[h_i,h_j]=0,\ \ [h_i,e_{j}] &=& \sign(j)C_{ij} e_j,\ \ [e_i,e_{-i}] =\sign(i) h_i,\nonumber
\\
({\rm Ad}\ e_i)^{1-C_{ij}}e_j &=& 0 \mbox{ for } i+j\neq 0
\eeq
The dual set $\{h^i\}$ is generated by $h_i=\sum_{j\in\Pi}C_{ij}h^j$. And then
\beq\label{relx1}
[h^i,h^j]=0,\ \ [h^i,e_{j}] &=& \sign(j)\delta_i^j e_j,\ \ [e_i,e_{-i}] =\sign(i) C_{ij}h^j,\nonumber
\\
({\rm Ad }\ e_i)^{1-C_{ij}}e_j &=& 0 \mbox{ for } i+j\neq 0\label{what}
\eeq
For any $i \in \Pi\cup{\bar\Pi}$, the group element can be expressed by $E_i = \exp(e_i)$ and a one-parameter subgroup is given by $H_i(z)=\exp(\log z\cdot h^{i})$.

Taking the $\mathfrak{g}=sl_N$ as example, the Cartan matrix is
\beq
\label{CslN1}
C_{ij}= (\alpha_i,\alpha_j)= 2\delta_{ij} - \delta_{i+1,j}-\delta_{i,j+1}\nonumber
\eeq
for the positive simple roots $\alpha_i\in\Pi$, $i,j=1,\ldots,{\rm rank}\ G=N-1$.

For generic Lie groups,
\beq\label{Cconstant}
C_{ij} = \langle\alpha_i,\alpha_j\rangle= {2\over (\alpha_i,\alpha_i)}(\alpha_i,\alpha_j)
= {1\over d_i}(\alpha_i,\alpha_j)
\eeq
and Cartan matrix is not symmetric for the non-simply-laced cases. In those cases, roots have different lengths and
$d_i = {1\over 2}(\alpha_i,\alpha_i)$ are not unities. Instead
of the Cartan matrix, for the non-simply-laced groups, defined as
\be\label{Cothergroup}
C_{ij} = d_iB_{ij},\ \ B_{ij} = {1\over d_id_j}(\alpha_i,\alpha_j),\ \  i,j\in\Pi
\ee
The dual vectors
\beq\label{daulvector}
\langle\alpha_i,\mu_j\rangle=\delta_{ij}
\eeq
are the highest weights of the fundamental representations. In this paper, we just only consider $sl(N)$ case. For other Lie groups, one need introduce special boundary conditions to find corresponding kink lattices which will be studied in the future.


\begin{thebibliography}{99}

\bibitem{Manton:2004tk}
  N.~S.~Manton and P.~Sutcliffe,
  ``Topological solitons,''
  (Cambridge University Press, Cambridge, England, 2004)

\bibitem{Vachaspati:2006zz}
  T.~Vachaspati,
  ``Kinks and domain walls: An introduction to classical and quantum solitons,''
  (Cambridge University Press, Cambridge, England, 2006)



  \bibitem{Dorey:2011yw}
  P.~Dorey, K.~Mersh, T.~Romanczukiewicz and Y.~Shnir,
  ``Kink-antikink collisions in the $\phi^6$ model,''
  Phys.\ Rev.\ Lett.\  {\bf 107}, 091602 (2011)
  [arXiv:1101.5951 [hep-th]].

\bibitem{Gani:2015cda}
  V.~A.~Gani, V.~Lensky and M.~A.~Lizunova,
  ``Kink excitation spectra in the (1+1)-dimensional $\varphi^8$ model,''
  JHEP {\bf 1508}, 147 (2015)
  [arXiv:1506.02313 [hep-th]].


\bibitem{Shifman:2004dr}
  M.~Shifman and A.~Yung,
  ``Non Abelian string junctions as confined monopoles,''
  Phys.\ Rev.\ D {\bf 70}, 045004 (2004)
  [hep-th/0403149].

\bibitem{Eto:2011cv}
  M.~Eto, T.~Fujimori, S.~B.~Gudnason, Y.~Jiang, K.~Konishi, M.~Nitta and K.~Ohashi,
  ``Vortices and Monopoles in Mass-deformed SO and USp Gauge Theories,''
  JHEP {\bf 1112}, 017 (2011)
  [arXiv:1108.6124 [hep-th]].


  \bibitem{AlonsoIzquierdo:2008rk}
  A.~Alonso-Izquierdo, M.~A.~Gonzalez Leon and J.~Mateos Guilarte,
  Phys.\ Rev.\ Lett.\  {\bf 101}, 131602 (2008)
  [arXiv:0808.3052 [hep-th]].

\bibitem{Harland:2009mf}
  D.~Harland,
  ``Kinks, chains, and loop groups in the $CP^N$ sigma models,''
  J.\ Math.\ Phys.\  {\bf 50}, 122902 (2009)
  [arXiv:0902.2303 [hep-th]].

\bibitem{Arai:2014hda}
  M.~Arai, F.~Blaschke, M.~Eto and N.~Sakai,
  ``Dynamics of slender monopoles and anti-monopoles in non-Abelian superconductor,''
  JHEP {\bf 1409}, 172 (2014)
  [arXiv:1407.2332 [hep-th]].


  \bibitem{toda}
Morikazu Toda.
\newblock Vibration of a chain with nonlinear interaction, J. Phys. Soc. Jpn. 22(2):431-436, 1967.


\bibitem{damianou90}
Pantelis~A. Damianou.
\newblock Master symmetries and {$R$}-matrices for the {T}oda lattice.
\newblock { Lett. Math. Phys.}, 20(2):101--112, 1990.

\bibitem{damianou91}
Pantelis~A. Damianou.
\newblock The {V}olterra model and its relation to the Toda lattice.
\newblock { Phys. Lett. A}, 155(2-3):126--132, 1991.

\bibitem{damianou93}
Pantelis~A. Damianou.
\newblock Symmetries of {T}oda equations.
\newblock { J. Phys. A}, 26(15):3791--3796, 1993.


\bibitem{olper} M.Olshanetsky and A.Perelomov,
Classical integrable finite-dimensional systems related to Lie algebras,
Phys.Peps., {\bf 71} (1981) 313;

\bibitem{olper1} M.Olshanetsky and A.Perelomov,
Quantum integrable systems related to Lie algebras,
Phys. Rep., {\bf 94} (1983) 6

\bibitem{olper2}
D. Kazhdan, B. Kostant and S. Sternberg,
Hamiltonian group action and dynamical systems of Calogero type,
Comm. on Pure and Appl. Math., Vol.XXXI (1978) 481-507

\bibitem{Marshakov:2012kv}
  A.~Marshakov,
  ``Lie Groups, Cluster Variables and Integrable Systems,''
  J.\ Geom.\ Phys.\  {\bf 67}, 16 (2013)
  [arXiv:1207.1869 [hep-th]].


\bibitem{Fock:2014ifa}
  V.~V.~Fock and A.~Marshakov,
  ``Loop groups, Clusters, Dimers and Integrable systems,''
  arXiv:1401.1606 [math.AG].

\bibitem{Kruglinskaya:2014pza}
  O.~Kruglinskaya and A.~Marshakov,
  ``On Lie Groups and Toda Lattices,''
  J.\ Phys.\ A {\bf 48}, no. 12, 125201 (2015)
  [arXiv:1404.6507 [hep-th]].


\bibitem{Fock:1997aia}
  V.~Fock and A.~Marshakov,
  ``A note on quantum groups and relativistic Toda theory,''
  Nucl.\ Phys.\ Proc.\ Suppl.\  {\bf 56B}, 208 (1997).




\bibitem{damianou02}
Pantelis~A. Damianou and Rui Loja~Fernandes,
\newblock From the {T}oda lattice to the {V}olterra lattice and back.
\newblock { Rep. Math. Phys.}, 50(3):361--378, 2002.


\bibitem{flaschka2}
H.~Flaschka.
\newblock On the {T}oda lattice. {II}. {I}nverse-scattering solution.
\newblock { Progr. Theoret. Phys.}, 51:703--716, 1974.


\bibitem{flaschka}
H.~Flaschka.
\newblock The {T}oda lattice. {I}. {E}xistence of integrals.
\newblock { Phys. Rev. B (3)}, 9:1924--1925, 1974.

\bibitem{henon}
M.~H{\'e}non.
\newblock Integrals of the {T}oda lattice.
\newblock { Phys. Rev. B (3)}, 9:1921--1923, 1974.


\bibitem{manakov}
S.~V. Manakov.
\newblock Complete integrability and stochastization of discrete dynamical
  systems.
\newblock {\em \v Z. \`Eksper. Teoret. Fiz.}, 67(2):543--555, 1974.



\bibitem{moser74}
J{\"u}rgen Moser.
\newblock Finitely many mass points on the line under the influence of an
  exponential potential--an integrable system.
\newblock In {\em Dynamical systems, theory and applications ({R}encontres,
  {B}attelle {R}es. {I}nst., {S}eattle, {W}ash., 1974)}, pages 467--497.
  Lecture Notes in Phys., Vol. 38. Springer, Berlin, 1975.





\bibitem{Gani:2014gxa}
  V.~A.~Gani, A.~E.~Kudryavtsev and M.~A.~Lizunova,
  ``Kink interactions in the (1+1)-dimensional $\phi^6$ model,''
  Phys.\ Rev.\ D {\bf 89}, no. 12, 125009 (2014)
  [arXiv:1402.5903 [hep-th]].






\bibitem{Bo} O.~I.~Bogoyavlensky,  On perturbation of the periodic
Toda lattices, Commun. Math. Phys. {\bf 51} (1976), 201--209.

  \bibitem{GKMMM}
A .Gorsky, I. Krichever, A. Marshakov, A. Mironov and A. Morozov,
Integrability and exact Seiberg-Witten solution,
Phys. Lett., {\bf B} 355 (1995) 466-477


\bibitem{6}
T. Nakatsu and K. Takasaki,
Whitham-Toda hierarchy and N=2
supersymmetric Yang-Mills theory,
Mod.Phys.Lett., {\bf A11} (1996) 157-168

\bibitem{61}
T. Eguchi and S. Yang,
Prepotentials of N=2 supersymmetric gauge theories and
soliton equations,
Mod. Phys. Lett., {\bf A11} (1996) 131-138


\bibitem{MW}
E. Martinec and N. Warner,
Integrable systems  and supersymmetric
Yang-Mills theory,
Nucl.Phys., {\bf 459} (1996) 97


\bibitem{Jiang:2016a}
Y. Jiang, J. Liu, The kink lattice is superintegrable, in preparation


\bibitem{Gani:2016txo}
  V.~A.~Gani, M.~A.~Lizunova and R.~V.~Radomskiy,
  ``Scalar triplet on a domain wall,''
  J.\ Phys.\ Conf.\ Ser.\  {\bf 675}, no. 1, 012020 (2016)
  [arXiv:1602.04446 [hep-th]].

\bibitem{Gudnason:2010rm}
  S.~B.~Gudnason, Y.~Jiang and K.~Konishi,
  ``Non-Abelian vortex dynamics: Effective world-sheet action,''
  JHEP {\bf 1008}, 012 (2010)
  [arXiv:1007.2116 [hep-th]].

  \bibitem{Eto:2015uqa}
  M.~Eto and M.~Nitta,
  ``Non-Abelian Sine-Gordon Solitons: Correspondence between $SU(N)$ Skyrmions and ${\mathbb C}P^{N-1}$ Lumps,''
  Phys.\ Rev.\ D {\bf 91}, no. 8, 085044 (2015)
  [arXiv:1501.07038 [hep-th]].

\bibitem{Nitta:2014rxa}
  M.~Nitta,
  ``Non-Abelian Sine-Gordon Solitons,''
  Nucl.\ Phys.\ B {\bf 895}, 288 (2015)
  [arXiv:1412.8276 [hep-th]].


  \bibitem{Ferreira:2010gh}
  L.~A.~Ferreira and W.~J.~Zakrzewski,
  ``The concept of quasi-integrability: a concrete example,''
  JHEP {\bf 1105}, 130 (2011)
  doi:10.1007/JHEP05(2011)130
  [arXiv:1011.2176 [hep-th]].

\bibitem{Ferreira:2016ubj}
  L.~A.~Ferreira, P.~Klimas and W.~J.~Zakrzewski,
  ``Quasi-integrable deformations of the $SU(3)$ Affine Toda Theory,''
  JHEP {\bf 1605}, 065 (2016)
  doi:10.1007/JHEP05(2016)065
  [arXiv:1602.02003 [hep-th]].


\end{thebibliography}
\end{document}